# Bayesian approach to extreme-value statistics based on conditional maximum-entropy method

Sumiyoshi Abe [1,2,3]

[1] Physics Division, College of Information Science and Engineering,
  Huaqiao University, Xiamen 361021, China

[2] Department of Physical Engineering, Mie University, Mie 514-8507, Japan

[3] Institute of Physics, Kazan Federal University, Kazan 420008, Russia

**Abstract**.   Recently, the conditional maximum-entropy method (abbreviated as C-MaxEnt) has been proposed for selecting priors in Bayesian statistics in a very simple way. Here, it is examined for extreme-value statistics. For the Weibull type as an explicit example, it is shown how C-MaxEnt can give rise to a prior satisfying Jeffreys' rule.

Understanding catastrophic events is a major goal of science of complex systems/phenomena. Examples include megaquakes in seismicity, crashes of financial markets, floods and so forth. A relevant mathematical concept may be extreme-value statistics [1]. Although occurrences of extreme events are rare, systems are strongly characterized by them, actually. It is noted that no big data are available about them, in general. This fact makes the Bayesian approach [2] attractive.

In the Bayesian inference scheme under the circumstance of limited data and/or uncertain knowledge, a key issue is how to select a prior. A lot of investigations have been devoted to this problem in the literature. From the viewpoint of physics, the efforts made by Jeffreys [3] and Jaynes [4] are obviously of particular interest since they base their selection rule upon the invariance principle in "information geometry", which somewhat reminds one of the measure of integration over spacetime in general relativity. In probability theory, however, a distribution is not invariant but accompanied by the Jacobian factor associated with a transformation of variables.

In a recent paper [5], we have formulated a new method of selecting a prior, which is referred to as the conditional maximum-entropy method and is abbreviated as C-MaxEnt. This method has originally been motivated by nonequilibrium statistical mechanics of complex systems governed by hierarchical dynamics with largely-separated time scales [6-8]. It enables one to express in a succinct way a prior probability merely in terms of a likelihood. It has been found [5] that C-MaxEnt can reproduce the classical results known for the scale and location variables and predicts an intriguing crossover-type (i.e., double power-law) prior for the Poissonian parameter.

In this article, we would like to examine C-MaxEnt for extreme-value statistics. In particular, we analyze Weibull statistics as an explicit example, in detail. We show how C-MaxEnt gives rise to a prior that is consistent with Jeffreys' rule.

Let us start our discussion with an explanation of the procedure of C-MaxEnt. A crucial point of importance is to identify a *conjugate pair of variables* (which should not be confused with conjugate priors). For this purpose, consider a couple of random variables, $X$ and $\Theta$, the realizations of which are denoted by $x$ and $\theta$, respectively. Each of them can be either continuous or discrete, but henceforth both of them are taken to be continuous. The joint probability distribution $P(x,\theta)$ is factorized as follows:

$$P(x,\theta) = p(x|\theta) f(\theta), \tag{1}$$

where $p(x|\theta)$ is the conditional probability distribution, i.e., a likelihood, and $f(\theta)$ is the marginal probability distribution, i.e., a prior. $(X, \Theta)$ is referred to as a conjugate pair, if the conditional variance of $X$ behaves as

$$(\Delta X)^2 \propto \frac{1}{\theta^2}. \tag{2}$$

If $p(x|\theta)$ does not have the second moment like the Lévy distribution, then the scaling property or the half width can be used instead of the variance [5]. In the case when there are more variables or odd number of variables, conjugate pairs may not be identified for some. In such a case, it is necessary to find conjugate pairs as many as possible (for details, see Ref. [5]). In view of Eq. (2), the following measure is defined:

$$d\mu = \left(\frac{dx}{k}\right)\left(\frac{d\theta}{l}\right). \tag{3}$$

Here, $k$ and $l$ are coarse-graining factors having the dimensions of $X$ and $\Theta$. For a continuous random variable, such a factor has to be introduced in order for the differential entropy to be definable. For our purpose, however, it does not actually play any role. Therefore, henceforth these factors are simply set equal to unity. Accordingly, the random variables are regarded to be dimensionless. Thus, the differential entropy is defined by

$$S[X, \Theta] = -\int dx\, d\theta\, P(x, \theta) \ln P(x, \theta). \tag{5}$$

From the physical viewpoint, Eq. (1) implies existence of the hierarchical structure in the "dynamics" of $X$ and $\Theta$: $X$ is a fast variable, whereas $\Theta$ is a slow one. This becomes manifest if Eq. (1) is used for rewriting the joint entropy in Eq. (5) as follows:

$$S[X, \Theta] = S[X | \Theta] + S[\Theta], \tag{5}$$

where $S[X|\Theta]$ and $S[\Theta]$ are the conditional and marginal entropies defined by

$$S[X|\Theta] = \int d\theta\, f(\theta)\, S[X|\theta], \qquad S[X|\theta] \equiv -\int dx\, p(x|\theta) \ln p(x|\theta), \tag{6}$$

$$S[\Theta] = -\int d\theta\, f(\theta) \ln f(\theta), \qquad (7)$$

respectively. It is noted that in Eq. (6) the fast variable is in fact integrated out (i.e., eliminated) first and then $S[X|\theta]$ as a function of $\theta$ is averaged afterward. This is in accordance with the way of treating hierarchical dynamics with largely-separated time scales in physics [8]. Actually, Eq. (5) is nothing but Khinchin's second axiom for the Shannon entropy in the discrete case [9].

C-MaxEnt for selecting a noninformative prior is now formulated as follows:

$$\delta_f \left\{ S[X, \Theta] - \lambda \int d\theta\, f(\theta) \right\} = 0. \qquad (8)$$

That is, the joint entropy is maximized only with respect to the prior, i.e., the slow degree of freedom, after elimination of the fast one. $\lambda$ stands for the Lagrange multiplier associated with the constraint on "normalization" condition on $f(\theta)$. Noninformativeness means that no constraints other than this constraint are imposed. In Bayesian statistics, $f(\theta)$ is often an improper prior, and therefore its integral over the whole range tends to diverge. This is why we do not specify in Eq. (8) the value of the integral of the prior (although in any way such a value turns out to be irrelevant). Using Eqs. (5)-(7) in Eq. (8), we obtain the following main result:

$$f(\theta) \propto \exp\{S[X|\theta]\}. \tag{9}$$

This shows how a noninformative prior can be expressed in terms only of a likelihood.

Now, let us apply C-MaxEnt to the problem of selecting a prior for extreme-value statistics [1]. Let us take $N$ i.i.d. random variables, $X_1$, $X_2$, ..., $X_N$, and consider the quantity, $(M_N - m_N)/\sigma_N$ with $m_N \in (-\infty, \infty)$ and $\sigma_N \in (0, \infty)$ being constants, where $M_N \equiv \max\{X_1, X_2, ..., X_N\}$. The Trinity Theorem states that there exist nontrivial domains of attraction in the limit $N \to \infty$, which lead to extreme-value statistics of the Gumbel, Weibull and Fréchet types. Since general discussions about priors for scale $\sigma_N$ and location $m_N$ have already been developed in Ref. [5], these are respectively set equal to unity and zero, for the sake of simplicity. Among the three types, here let us treat the one of Weibull as a typical explicit example. The Weibull distribution is given by

$$p(x|\alpha) = \frac{1}{\alpha}(-x)^{1/\alpha - 1} \exp\left[-(-x)^{1/\alpha}\right], \tag{10}$$

where

$$x \in (-\infty, 0], \qquad \alpha \in (0, \infty). \tag{11}$$

Our task is to perform Bayesian inference of the value of the parameter $\alpha$ through its prior by the use of C-MaxEnt, in which $p(x|\alpha)$ in Eq. (10) is regarded as the likelihood. Following the procedure of C-MaxEnt, let us find a conjugate pair of variables. The first and second moments are given by $\langle X \rangle = \int_{-\infty}^{0} dx\, x\, p(x|\alpha) = -\Gamma(\alpha+1)$ and $\langle X^2 \rangle = \Gamma(2\alpha+1)$, respectively, where $\Gamma(s)$ is the gamma function. So, the conditional variance reads

$$(\Delta X)^2 = \Gamma(2\alpha+1) - [\Gamma(\alpha+1)]^2. \tag{12}$$

This implies that the variable conjugate to $X$ is

$$\Theta = \left\{ \Gamma(2A+1) - [\Gamma(A+1)]^2 \right\}^{-1/2}, \tag{13}$$

where $A$ is the random variable whose realization is $\alpha$. $d\theta$ and $d\alpha$ are thus related to each other through the Jacobian factor as follows:

$$d\theta = d\alpha\, \frac{\psi(2\alpha+1)\,\Gamma(2\alpha+1) - \psi(\alpha+1)[\Gamma(\alpha+1)]^2}{\left\{ \Gamma(2\alpha+1) - [\Gamma(\alpha+1)]^2 \right\}^{3/2}}, \tag{14}$$

where $\psi(s)$ is the digamma function. On the other hand, $S[X|\alpha]$ defined in the way similar to Eq. (6) is calculated to be

$$S[X|\alpha) = \ln\alpha + \gamma(1-\alpha) + 1, \tag{15}$$

where $\gamma = 0.57721...$ is Euler's constant. Therefore, the prior is found to be

$$f(\alpha) \propto \alpha e^{-\gamma\alpha} \frac{\psi(2\alpha+1)\Gamma(2\alpha+1) - \psi(\alpha+1)[\Gamma(\alpha+1)]^2}{\left\{\Gamma(2\alpha+1) - [\Gamma(\alpha+1)]^2\right\}^{3/2}}. \tag{16}$$

The plot of $f(\alpha)$ is presented in figure 1.

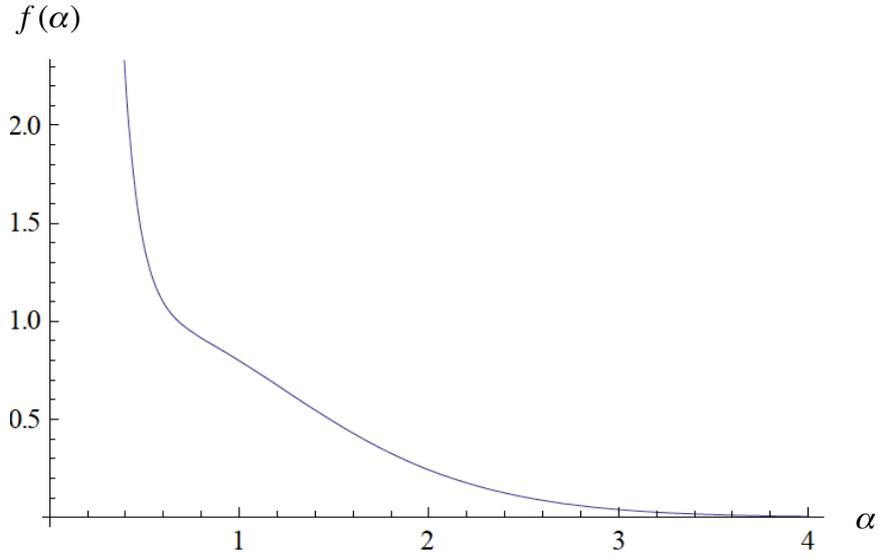

**Figure 1.** Plot of $f(\alpha)$ in Eq. (16) with respect to $\alpha$, where the proportionality constant is set equal to unity. All quantities are dimensionless.

The asymptotic behaviors of $f(\alpha)$ are analytically evaluated as follows:

$$f(\alpha) \sim \frac{1}{\alpha} \ (\alpha \to 0+), \ \exp(-\alpha \ln \alpha) \ (\alpha \to \infty). \tag{17}$$

Note that the linear divergence in the limit $\alpha \to 0+$ is plausible in view of Jeffreys' rule [4].

In conclusion, we have applied C-MaxEnt to the problem of selecting a prior in Bayesian approach to extreme-value statistics. Taking the Weibull type as an explicit example, we have shown how C-MaxEnt can yield a prior that is reasonable in view of Jeffreys' rule.

**Acknowledgments**

This work has been supported in part by grants from National Natural Science Foundation of China (No. 11775084) and Grant-in-Aid for Scientific Research from the Japan Society for the Promotion of Science (No. 26400391 and No. 16K05484), and by the Program of Competitive Growth of Kazan Federal University from the Ministry of Education and Science of the Russian Federation.